\documentclass[justified]{tufte-handout}

\usepackage{color}
 
\newcommand{\red}{\textcolor[rgb]{0.75,0.00,0.00} }

\title[\uppercase{N}eural wave interference and intrinsic tuning]{\bf 
	\red{Neural wave interference and intrinsic tuning \vspace{3pt} 
	 \\ in distributed excitatory-inhibitory networks  } }

\date{ }

\usepackage{graphicx} 
\setkeys{Gin}{width=\linewidth,totalheight=\textheight,keepaspectratio}
\graphicspath{{graphics/}} 
\usepackage{amsmath}  
\usepackage{booktabs} 
\usepackage{units}    
\usepackage{multicol} 
\usepackage{lipsum}   
\usepackage{fancyvrb} 
\fvset{fontsize=\normalsize}

\begin{document}

\maketitle
\vspace{-6pt}
\begin{fullwidth}
\noindent 
Sergei Gepshtein,$^1$ Ambarish S. Pawar,$^1$ Sergey Savel'ev,$^2$ Thomas D. Albright$^1$ 

{\it 
\vspace{10pt}
\noindent 
$^1$ Center for Neurobiology of Vision, Salk Institute for Biological Studies
\\ 10010 North Torrey Pines Road, La Jolla, CA 92037, USA

\vspace{5pt}
\noindent 
$^2$ Department of Physics, Loughborough University \\ Leicestershire, LE11 3TU, United Kingdom
}

\vspace{.2in}

\noindent 
ABSTRACT. 
We developed a model of cortical computation that implements key features of cortical circuitry and is capable of describing propagation of neural signals between cortical locations in response to spatially distributed  stimuli. 
The model is based on the canonical neural circuit that consists of excitatory and inhibitory cells interacting through reciprocal connections, with recurrent feedback.  
The canonical circuit is used as a node in a distributed network with nearest neighbor coupling between the nodes. 
We find that this system is characterized by intrinsic preference for spatial frequency. 
The value of preferred frequency depends on the relative weights of excitatory and inhibitory connections between cells. 
This balance of excitation and inhibition changes as stimulus contrast  increases, which is why intrinsic spatial frequency is predicted to change with contrast in a manner determined by stimulus temporal frequency. 
The dynamics of network preference is consistent with properties of the cortical area MT in alert macaque~monkeys.
\end{fullwidth}

\tableofcontents 
\listoffigures

\newpage
\section{Introduction}

\noindent  
We sought to develop a framework for modeling visual cortical processes satisfying a number of requirements:

\begin{description}
\item Biological realism:~Include key features of cortical circuity that consists of excitatory and inhibitory neurons connected  reciprocally and affected by recurrent feedback.

\item Distributed dynamics:~Allow for propagation of signals between cortical locations, in order to predict responses to stimuli distributed spatially and temporally. 

\item Mathematical tractability:~Essential properties of dynamics should be tractable by methods of analysis and not numerical simulation alone.  

\end{description}

\begin{figure*}[t!]
	\centering
	\includegraphics[width=.95\linewidth]{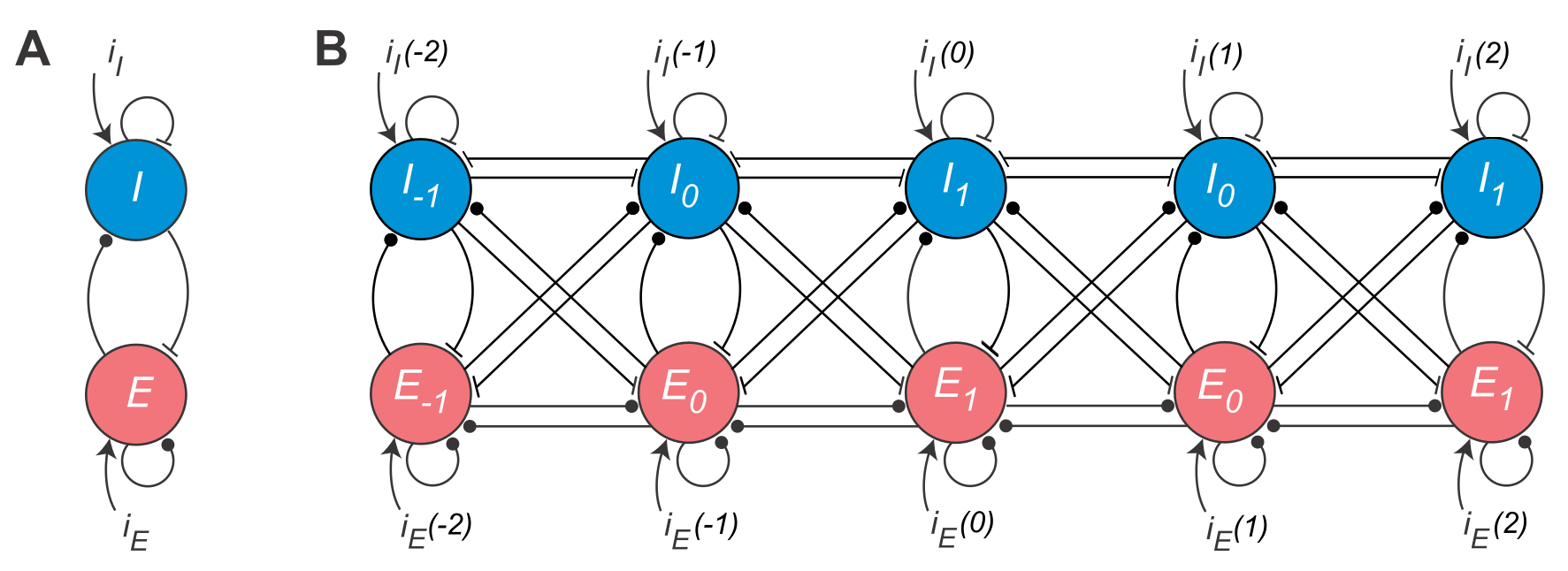}
	{
		\caption[The canonical circuit][0.28in]{  
			The canonical circuit.  
			\vspace{3pt} \newline 
			A.~The canonical Wilson-Cowan circuit contains one excitatory unit (\emph{E}) and one inhibitory unit (\emph{I}), each with recurrent feedback, and connected reciprocally (i.e., with the excitatory cell connected to the inhibitory cell, and vice versa).~In the figure, the lines ending with small filled circles and with T-junctions represent, respectively, excitatory and inhibitory connections between the cells.
			\vspace{3pt} \newline 
			B.~The canonical circuits are arranged as ``nodes" in a chain with nearest-neighbor coupling. Indices $l$ of the nodes (notated as $E_l$ and $I_l$) indicate node locations in the chain. Currents $i_E(l)$ and $i_I(l)$ are the inputs into each node generated by the stimulus.		
		} 
	}
	\label{fig:chain}
	\setfloatalignment{t}
\end{figure*}

\vspace{0in} \noindent  Accordingly, we developed a spatially distributed model based on the canonical excitatory-inhibitory circuit introduced by \citet{wilson1972excitatory,wilson_cowan1973}. 
The canonical circuit is illustrated in Figure~\ref{fig:chain}A. 
It consists of two units ("cells"): excitatory and inhibitory ($E$ and $I$), connected reciprocally and recurrently. 
Previous studies of this circuit suggested that it can serve as a universal motif for dynamic models of cortical neural networks \citep[e.g.,][]{tsodyks1997paradoxical,ozeki2009inhibitory,ahmadian2013analysis,jadi2014regulating,rubin2015stabilized}.\sidenote[][-40pt]{Of particular note is the discovery by  \citet{tsodyks1997paradoxical} that inhibitory-inhibitory connections are essential for providing stability in the otherwise unstable neural networks with recurrent excitation.~This work showed that the  architecture with recurrent excitation and inhibitory-inhibitory stabilization allows for intriguing variety of dynamic regimes in the network.}  

In our spatially distributed model, the canonical circuit serves as a network motif: a repetitive ``node" in a larger network (Figure~\ref{fig:chain}B). 
This network architecture is a conservative generalization of the canonical circuit: from the spatially local system of Wilson and~Cowan to a spatially distributed system suitable for modeling neural response to stimuli distributed both spatially and temporally \citep{SavelievGepshtein2014entropy}. 
That is, the excitatory and inhibitory units of each node are each connected to both the excitatory and inhibitory units of the neighboring nodes, which is a form of connectivity least removed from the canonical model. 

We used two analytical approaches. 
The first approach was inspired by the method of solving {\it linear} differential equations using Green's functions \citep{riley2006mathematical}, commonly used in the analysis of mechanical and electrical systems \citep{challis2003green}. 
A Green's function is a response of the system to impulse activation: a signal localized spatially or temporally. 
Having found the Green's function of the system, responses to distributed  activation are represented by convolving the  Green's function with the distributed  activation. 
This approach revealed that the spatially distributed canonical circuit has a number of interesting emergent properties.
One of these properties is the intrinsic preference of the system to the spatial frequency of stimulation. 
The system responds to spatially periodic stimuli of a certain spatial frequency more vigorously than other frequencies. 

In the second approach, we investigated how this intrinsic stimulus preference chan\-ges at high stimulus contrasts, where the system becomes {\it nonlinear}. 
We found that  the intrinsic spatial frequency of the system strongly depends on the relative weights of excitation and inhibition in cell connections.  
We also found that this balance of excitatory and inhibitory weights changes as luminance contrast of stimuli increases, which is why system preferences are predicted to change with contrast, in a manner determined by stimulus temporal frequency.  

The dynamics of system preference revealed in this study departs from the traditional view of neuronal selectivity, but our predictions happen to be strikingly similar to the dynamics of neuronal preferences discovered recently in a study of the Middle Temporal (MT) cortical area of alert macaque monkeys (\citealt{Pawar_etal_2017SfN,Pawar_etal_2018cosyne}; Pawar et al., in preparation).

\section{Approach}

Consider a system of Wilson-Cowan equations for a chain of "nodes" each containing an excitatory cell ("$E$") and an inhibitory cell  ("$I$"), as shown in Figure~\ref{fig:chain}:
{\small
\begin{eqnarray}
\tau_E\frac{dr_E(l)}{dt}=-r_E+g_E({\cal W}_E)\nonumber, \\
\frac{dr_I(l)}{dt}=-r_I+g_I({\cal W}_I).
\end{eqnarray}
}
The variables $r_E$ and $r_I$ represent the firing rates of the excitatory and inhibitory cells, $\tau_E$ represents the relaxation time of excitation (in units of the relaxation time of inhibition),  $l$ is the node index in the chain,  and $g_E$, $g_I$ are sigmoid functions.  ${\cal W}_E$ and ${\cal W}_I$ are positive weights:  
\[ {\cal W}_E= 
\Bigl[w_{EE}r_E(l)+{\tilde w}_{EE}r_E(l+1)+{\tilde w}_{EE}r_E(l-1)\Bigr] -
\Bigl[w_{EI}r_I(l)+{\tilde w}_{EI}r_I(l+1)+{\tilde w}_{EI}r_I(l-1)\Bigr]        + i_E(l,t), \]
\[ {\cal W}_I=
\Bigl[w_{IE}r_E(l)+{\tilde w}_{IE}r_E(l+1)+{\tilde w}_{IE}r_E(l-1)\Bigr] -
\Bigl[w_{II}r_I(l)+{\tilde w}_{II}r_I(l+1)+{\tilde w}_{II}r_I(l-1)\Bigr]        + i_I(l,t), \]
where $w$ and ${\tilde w}$ represent the weights of connections respectively within and between the nodes. 
Assuming that stimulus inputs $i_E, i_I$ and cell responses $r_E, r_I$ vary slowly on the scale of inter-node distance, we replace the discrete node index $l$ by a continuous spatial variable $x$, to obtain: 
\[ r_E(l\pm 1)=r_E(l)\pm\frac{\partial r_E}{\partial x}+\frac{1}{2}\frac{\partial^2 r_E}{\partial x^2}, \] 
\[ r_I(l\pm 1)=r_I(l)\pm\frac{\partial r_I}{\partial x}+\frac{1}{2}\frac{\partial^2 r_I}{\partial x^2}. \] 
By expanding the inverse sigmoid  functions $g_{E}^{-1}$ and $g_{I}^{-1}$ in the Taylor series up to the third order, we obtain from the above the following pair of coupled partial differential equations:
{\small 
\begin{eqnarray}
K_{EE}r_E(x,t)+D_{EE}\frac{\partial^2 r_E}{\partial x^2}-K_{EI}r_I(x,t)-
D_{EI}\frac{\partial^2 r_I}{\partial x^2}+\alpha j(x,t)\nonumber \\
=g_E^{-1}\left(\frac{\partial r_E}{\partial t}+r_E\right)\approx
\left(\frac{\partial r_E}{\partial t}+r_E\right)+\beta_E\left(\frac{\partial r_E}{\partial t}+r_E\right)^2
+\gamma_E \left(\frac{\partial r_E}{\partial t}+r_E\right)^3, \nonumber \\
K_{IE}r_E(x,t)+D_{IE}\frac{\partial^2 r_E}{\partial x^2}-K_{II}r_I(x,t)-
D_{II}\frac{\partial^2 r_I}{\partial x^2}+(1-\alpha) j(x,t)\nonumber \\
=g_I^{-1}\left(\frac{\partial r_I}{\partial t}+r_I\right)\approx
\left(\frac{\partial r_I}{\partial t}+r_I\right)+\beta_I\left(\frac{\partial r_I}{\partial t}+r_I\right)^2
+\gamma_I \left(\frac{\partial r_I}{\partial t}+r_I\right)^3,
\label{eq:pde}
\end{eqnarray}
}
where  $K$ are the ``interaction constants" reflecting  the strengths of connections between the excitatory and inhibitory parts of the network ($K_{EE}=w_{EE}+2{\tilde w}_{EE}$, $K_{EI}=w_{EI}+2{\tilde w}_{EI}$, etc), and $D$ are the ``diffusion constants" reflecting  the strengths of connections between the nodes ($D_{EE}={\tilde w}_{EE}$, $D_{EI}={\tilde w}_{EI}$, etc) and thus are responsible for spatial propagation of excitatory and inhibitory influences through the network.
The Taylor expansion coefficients are $\beta$ (second order) and $\gamma$ (third order). 
Among these coefficients, $\gamma$ and $\beta$ can be different for excitatory ($\beta_E, \gamma_E$) and inhibitory ($\beta_I, \gamma_I$) cells; 
they define the degree of nonlinearity in the system.
Parameter $\alpha$ describes how the input current $i$ is divided between the excitatory and inhibitory cells. 
That is, in a system activated by the spatiotemporal  stimulus $j(x,t)$, the input currents are $i_E(x,t)=\alpha j(x,t)$ and $i_I(x,t)=(1-\alpha) j(x,t)$.

\section{Results}
	
\subsection{Point-source waves} 

First, consider how  the system responds to a "point" stimulus  $j(x,t)=\delta(x)$ that activates a singe node of the chain. 
This question requires that we solve system~(\ref{eq:pde}) with zero $j(x,t)$, separately for $x<0$ and $x>0$,
in linear approximation (where $\gamma_E=\gamma_I=\beta_E=\beta_I=0$).
For the analysis of system response to localized stimuli, we notate responses as $r_E(x)=G_E(x)$ and $r_I=G_I(x)$. 
We find the solutions separately for the left and right sides of  the stimulus and then match the solutions at $x=0$ so that:\sidenote[][-50pt]{
	Here we collect the auxiliary coefficients used henceforth:
	\vspace{.1in}
	\newline $\mu=D_{EI}D_{IE}-D_{EE}D_{II},$
	\newline $b=(1/2\mu)(D_{EI}K_{IE}+D_{IE}K_{EI} - $  
	\newline \hspace{.3in} $ - (K_{EE}-1)D_{II} -D_{EE}(K_{II}+1)),$
	\newline $d=((1-K_{EE})(1+K_{II})+$
	\newline \hspace{.96in} $+K_{EI}K_{IE})/\mu - b^2,$
	\newline $\Psi_I=\alpha(K_{II}+1)-(1-\alpha)K_{EI},$
	\newline $\Psi_E=\alpha K_{IE}-(1-\alpha)(K_{EE}-1),$ 
	\newline $\Phi_I=(1-\alpha)D_{EI}-\alpha D_{II},$ 
	\newline $\Phi_E=(1-\alpha)D_{EE}-\alpha D_{IE},$
	\newline $\eta_E=(3/4)(1-\alpha)\gamma_E,$ 
	\newline $\eta_I=(3/4)\alpha\gamma_I,$
	\newline $ \xi_E=(3/8)D_{II}\gamma_E/\mu,$
	\newline $ \xi_I=(3/8)D_{EE}\gamma_I/\mu,$
	\newline $ \sigma_E=(3/4)(K_{II}+1)\gamma_E/\mu+2b\xi_E,$ 
	\newline $ \sigma_I=(3/4)(K_{EE}-1)\gamma_I/\mu+2b\xi_I,$
	\newline $\Gamma_E=\frac{\sqrt{d}D_{EI}\Delta+\Lambda(K_{EI}-D_{EI}b)/\sqrt{d}}{D_{EI}(K_{EE}-1-D_{EE}b)-D_{EE}(K_{EI}-D_{EI}b)},$
	\newline $\Gamma_I=\frac{\sqrt{d}D_{EE}\Delta+\Lambda(K_{EE}-1-D_{EE}b)/\sqrt{d}}{D_{EI}(K_{EE}-1-D_{EE}b)-D_{EE}(K_{EI}-D_{EI}b)},$
	\newline $\Delta=D_{EI\Phi_E-D_{EE}\Phi_I},$
	\newline $\Lambda=(K_{EI}-D_{EI})\Phi_E-$
	\newline \hspace{.6in} $(K_{EE}-1-D_{EE}b)\Phi_I$.
	\label{foot:aux}}   
\begin{align}
G_E(+0) =& G_E(-0), \nonumber \\
G_I(+0) =& G_I(-0), \nonumber \\
\left.\frac{\partial G_E}{\partial x}\right|_{+0}-
\left.\frac{\partial G_E}{\partial x}\right|_{-0} =& -\frac{1}{\mu}\Phi_I, \nonumber \\ 
\left.\frac{\partial G_I}{\partial x}\right|_{+0}-
\left.\frac{\partial G_I}{\partial x}\right|_{-0} =& -\frac{1}{\mu}\Phi_E
\label{eq:conditions}
\end{align}
We then substitute the solutions for $x<0$ and $x>0$ into (\ref{eq:conditions}) and obtain:
\begin{eqnarray}
G_E=\frac{1}{2k_n\mu}e^{-\lambda |x|}\left(\Gamma_E \cos(k_nx)-\Phi_I{\rm sign}(x)\sin(k_nx) \right), \nonumber \\
G_I=\frac{1}{2k_n\mu}e^{-\lambda |x|}\left(\Gamma_I \cos(k_nx)-\Phi_E{\rm sign}(x)\sin(k_nx) \right),
\label{eq:stat_wave}
\end{eqnarray}
where ${\rm sign}(x)=x/|x|$ for $x\ne 0$ and ${\rm sign}(0)=0$, and where\sidenote[][10pt]{Notice that $k_n^2 = b$ for $d=0$ which will serve a useful reference for Equation~\ref{eq:summary_rewrite}.}
\begin{equation}
k_n^2={   \frac{ b+\sqrt{b^2+d}}{2} }, \quad
\lambda= \frac{\sqrt{d}} {2k_n}.
\end{equation}

\begin{marginfigure}[-70pt]
	\includegraphics[width=1.0\linewidth]{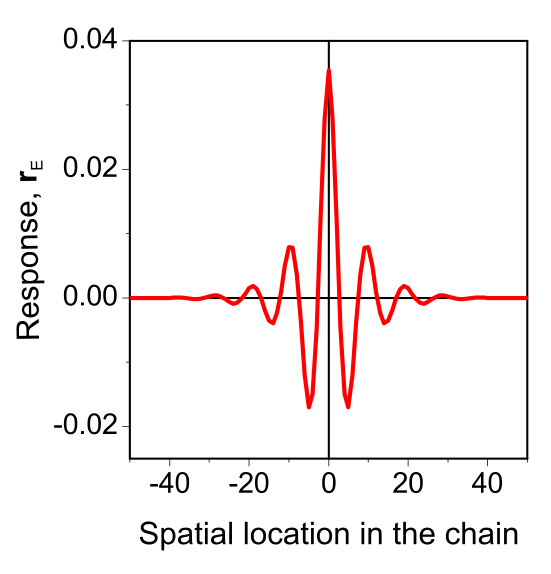} 
	\caption[Response of the distributed circuit to a point stimulus]{
		Response of the distributed circuit to a small spatial stimulus (``point stimulus") that activates a single node of the circuit. 
		\vspace{3pt} \newline 
		The point activation propagates through the chain and forms a steady-state spatial pattern distributed across the chain. This pattern is an instance of neural wave and it has the form of spatially damped oscillation.
	} 
	\label{fig:green}
	\vspace{1in}
\end{marginfigure}

Equation~\ref{eq:stat_wave} determines a spatial oscillation (illustrated in~Figure~\ref{fig:green}) with the spatial frequency of $k_n$.
The period of spatial oscillation is
\begin{equation}
L = \frac{2 \pi}{k_n}.
\label{eq:spatial_period}
\end{equation}
The term $\lambda$ determines the rate of spatial decay of system response. 
For very small $\lambda$ ($\lambda\ll 1$ and thus $\sqrt{d}\ll b$), we have $k_n\approx \sqrt{b}$. 
Spatial frequency $k_n$ can be called the natural or {\it intrinsic} spatial frequency of the system.

For stimuli  more complex than a delta function, the general static solution in linear approximation can be written as:
\begin{eqnarray}
r_E=\int_{-\infty}^{\infty} j(x')G_E(x-x')dx',\nonumber\\
r_I=\int_{-\infty}^{\infty} j(x')G_I(x-x')dx'.
\end{eqnarray}
\noindent suggesting how our model is related to the {\it standard description} of visual neural mechanisms  in terms of {linear filters} \citep{CampbellRobson1968,koenderinkVanDoorn1987representation,devalois1988spatial,graham1989visual,watson2016pyramid}. 
To account for nonlinear properties of neural responses, the standard approach assumes that the stage of linear filters is followed by a separate nonlinear stage.

In contrast to the standard picture, a single mechanism is used in our approach to account for both linear and nonlinear regimes, without postulating a separate nonlinear process.  
Indeed, by analogy of our system with a system of coupled oscillators (e.g., \S28 in \citealp{Landau1969mechanics}), it is expected that the intrinsic frequency of our system should change with contrast, as we show just below.

\subsection{Distributed periodic stimulation}

Next we studied system responses to higher contrasts, where nonlinear effects become significant,  and so we should consider terms $\gamma_E, \gamma_I, \beta_E, \beta_I$.
In writing the solutions of (\ref{eq:pde}) we have found, however, that our main results  (\ref{eq:iter}-\ref{eq:summary_rewrite}) about the interaction of stimulus contrast and intrinsic spatial frequency of the circuit do not change qualitatively when terms $\beta$ are ignored. 
We therefore present results of this analysis in a simplified form, omitting the quadratic terms in Taylor expansions. 
And to obtain steady-state solutions, in this section we also ignore time derivatives and  the time course of stimulus (which we do consider in the next section).
We derive system response to periodic stimuli of the form:
\[j(x)=j_0\cos kx,\]
where $k$ is spatial frequency. 
We find the solutions in the form
\begin{eqnarray}
r_E(x,t)=\mathcal{E}\cos kx
\nonumber \\
r_I(x,t)=\mathcal{I}\cos kx,
\label{eq:soluitions1}
\end{eqnarray} 
where constants 
$\mathcal{E}$       
and $\mathcal{I}$ 
represent the amplitudes of excitatory and inhibitory activations. 
While keeping only the main harmonics in the solution, we obtain:
{\small
\begin{eqnarray}
\left(K_{EE}-1-D_{EE}k^2-\frac{3}{4}\gamma_E\left(\mathcal{E}\right)^2\right)\mathcal{E}+\left(D_{EI}k^2-K_{EI}\right)\mathcal{I}=-\alpha j_0, \nonumber \\
\left(K_{IE}-D_{IE}k^2\right)\mathcal{E}+\left(k^2D_{II}-K_{II}-1-\frac{3}{4}\gamma_I\left(\mathcal{I}\right)^2\right)\mathcal{I}=-(1-\alpha)j_0.
\label{eq:solution3}
\end{eqnarray}
}

\begin{figure*}[t!]
	\includegraphics[width=.85\linewidth]{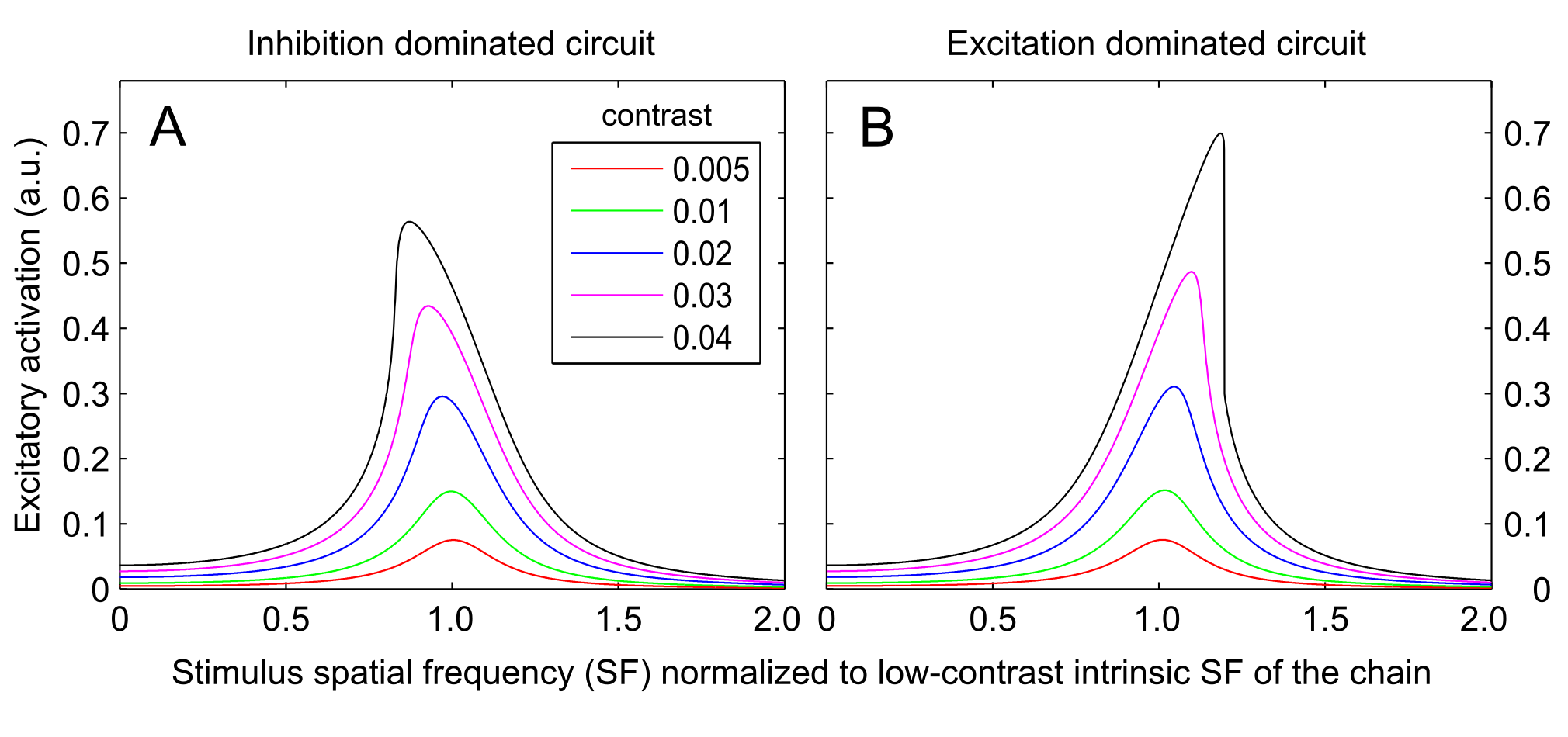}
	{
		\caption[Response of the distributed circuit to increasing contrast][.25in]{Response of the distributed circuit to increasing contrast.  
			\vspace{3pt} \newline 
			The curves in each panel are iterative solutions of (\ref{eq:iter}--\ref{eq:iter2}). 
			Each curve represents activation of one excitatory cell ($E_0$ in Figure~\ref{fig:chain}B) at one stimulus contrast, plotted as a function of stimulus spatial frequency. 
			Stimulus spatial frequency is normalized to the resonant spatial frequency of the network measured at the lowest tested contrast (the red curves).
			\vspace{3pt} \newline 
			In general, increasing stimulus contrast reveals different magnitudes of the resonant spatial frequency: falling resonant frequency in a system dominated by inhibition~(A); rising resonant frequency in a system dominated by excitation~(B). 		
		} 
	}
	\label{fig:slices}
	\setfloatalignment{t}
\end{figure*}

The system (\ref{eq:solution3}) can be solved iteratively: 
\begin{eqnarray}
\mathcal{I}_{n+1}= C\frac{\Psi_E+\Phi_E k^2+\eta_I \mathcal{I}_{n}^2}{\left(k^2-b+\xi_I \mathcal{I}_n^2-\xi_E\mathcal{E}_n^2\right)^2+d+
\sigma_I \mathcal{I}_n^2-\sigma_E\mathcal{E}_n^2}
\label{eq:iter}
\nonumber, 
\\
\\
\mathcal{E}_{n+1}=C\frac{\Psi_I+\Phi_I k^2+\eta_E \mathcal{E}_{n}^2}{\left(k^2-b+\xi_I \mathcal{I}_n^2-\xi_E\mathcal{E}_n^2\right)^2+d+
\sigma_I \mathcal{I}_n^2-\sigma_E\mathcal{E}_n^2},
\label{eq:iter2}
\end{eqnarray}
where $C=j_0/\mu$ is the effective stimulus contrast.\sidenote[][-91pt]{As we show below, some of the coefficients used in the system (\ref{eq:iter}-\ref{eq:iter2}) have an immediate interpretation in terms of intrinsic preference of the neural chain. In particular, $b$ determines the location of the peak of selectivity at low stimulus contrast and zero temporal frequency (\ref{eq:summary}), and $d$ determines the degree of selectivity (i.e., the ``width" of tuning to spatial frequency).}
The new system~(\ref{eq:iter}-\ref{eq:iter2}) allows us to derive results of  the ($n+1$)th iterations  for $\mathcal{E}_{n+1}$ and $\mathcal{I}_{n+1}$ using results of the $n$th iterations for $\mathcal{E}_{n}$ and $\mathcal{I}_{n}$.  The iterative procedure converges for the values of coefficients described below; the results are plotted  in~Figure~\ref{fig:slices}.

Figure~\ref{fig:slices} is a plot of iterative solutions of system (\ref{eq:iter}-\ref{eq:iter2})~with the coefficients $(\xi_E,\xi_I)$ set to (1, 0.2) in panel~B and  (0.2, 1) in panel~C.  
Effective stimulus contrast is varied by setting $C = j_0/\mu$ to five values (0.005, 0.01, 0.02, 0.03, 0.04) for which the  responses of 100 iterations are plotted against the spatial frequency $k$, normalized to the resonance frequency, $k_{\rm res}$ at the lowest contrast (shown in red). 
The plots reveal that cell tuning shits with contrast: toward lower $k$ in the inhibition-dominated network ($\xi_I>\xi_E$, panel~B) and toward higher $k$ in the excitation-dominated network ($\xi_E>\xi_I$, panel~C). 
Other parameters are $b$=1, $d$=0.1, $\Psi_E$=$\Psi_I$=1, $\Phi_E$=$\Phi_I$=0.5, $\eta_E$=$\eta_I$=0.1, $\sigma_E$=$\sigma_I$=0.01. 

Notice that in the system (\ref{eq:iter}-\ref{eq:iter2}), $\sigma$ affects the bandwidth of response, and $\xi$ shifts the location of the maximum of $\mathcal{E}$ on the dimension of spatial frequency $k$. 
The magnitude of $k$ at which $\mathcal{E}$  reaches its maximum is the \emph{characteristic} (or \emph{resonant}) spatial frequency of the network, $k=k_{\rm res}$. 
Notice also that the maximum of $\mathcal{E}$ in this system is found near the minimum of the denominator of  (\ref{eq:iter}-\ref{eq:iter2}).
Conditions of this minimum can  be written explicitly: 
{
\begin{equation}
k_{\rm res}^2=b-\xi_I \, \mathcal{I}^2+\xi_{E} \, \mathcal{E}^2.
\label{eq:summary}
\end{equation}
}
Remarkably, the terms for $\mathcal{I}$ and $\mathcal{E}$ in (\ref{eq:summary}) have opposite signs. 
To~appreciate this result, suppose that the activation of excitatory and inhibitory cells within a node have similar magnitudes $\mathcal{A} = \mathcal{E} = \mathcal{I}$. 
We can therefore rewrite (\ref{eq:summary}) as
\begin{equation}
k_{\rm res}^2= b -  \mathcal{A}^2 (\xi_I -  \xi_E).
\label{eq:summary_rewrite}
\end{equation}
Recall that $b$, $\xi_I$, and $\xi_E$ are constants and that $\mathcal{A}$ is an increasing function of contrast. 
It follows that the second term of (\ref{eq:summary_rewrite}), which is $\mathcal{A}^2 (\xi_I - \xi_E)$, will increase with contrast when $\xi_I>\xi_E$ and the resonant frequency $k_{\rm res}$ will decrease~(Figure~\ref{fig:slices}A).  
And when $\xi_I<\xi_E$, the second term of (\ref{eq:summary_rewrite}) will decrease with contrast and  $k_{\rm res}$ will increase~(Figure~\ref{fig:slices}B). 

In summary, the network's intrinsic preference for stimulus spatial frequency ($k_{res}$) will generally change with stimulus contrast: 
$k_{res}$ will increase with contrast in a system dominated by excitation, and $k_{res}$ will decrease with contrast in a system dominated by inhibition.

\subsection{Distributed spatiotemporal stimulation}

Now we consider dynamic stimuli: drifting luminance gratings characterized in terms of both spatial and temporal frequencies, which we represent by 
\[j(x,t)=j_0\cos(\omega t - kx),\] 
where $\omega$ is temporal frequency. 
Here we concentrate on low luminance contrasts, where the nonlinear terms can be omitted. 
As we show below,  results of this analysis have a form that allows us to make predictions for the system behavior also at high luminance contrasts.

\begin{figure*}[t!]
	\includegraphics[width=.95\linewidth]{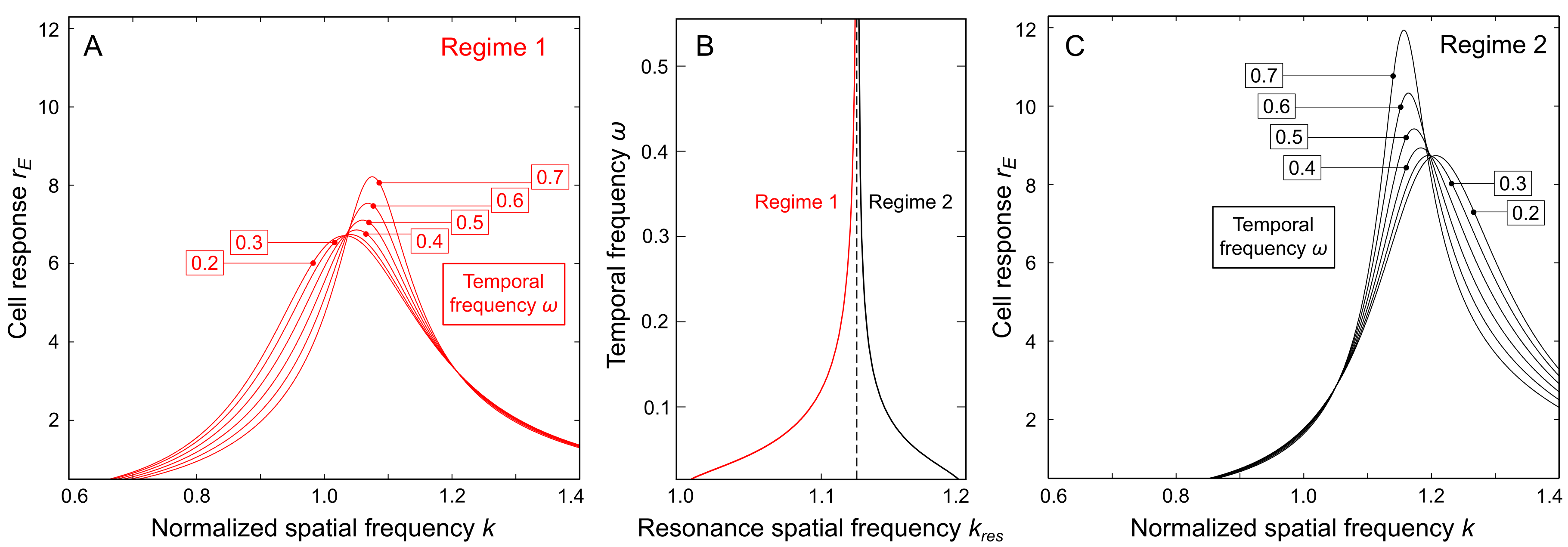} 
	\caption[Intrinsic spatial frequency depends on stimulus temporal frequency][18pt]{Intrinsic tuning to stimulus spatial frequency is predicted to depend on stimulus temporal frequency. 
		\vspace{3pt} \newline 
		A.~Response functions are plotted at different values of stimulus temporal frequency~$\omega$ when the system is in Regime~1. 
		The peak of each tuning function is the intrinsic spatial frequency of the system for the value of~$\omega$ displayed in the corresponding box.~Intrinsic spatial frequency $k_{res}$ increases with~$\omega$. 
		\vspace{3pt} \newline 
		B.~The plot at left, in red, describes Regime~1. 
		Here intrinsic spatial frequency of the system increases with~$\omega$, approaching a vertical asymptote from the left. 
		The plot at right, in black, describes Regime~2 (as in panel~C), in which intrinsic spatial frequency decreases with~$\omega$, approaching the same vertical asymptote from the right.
		\vspace{3pt} \newline 
		C.~Response functions are plotted as in panel~A for the system in Regime~2. 
		Here intrinsic spatial frequency decreases with~$\omega$.
	} 
	\label{fig:sf_tf}
	\setfloatalignment{t}
\end{figure*}

We therefore seek solutions of (\ref{eq:pde}) in the following form: 
\begin{eqnarray}
r_E(x,t)=\mathcal{E}_c\cos(\omega t - kx)+\mathcal{E}_s\sin(\omega t-kx)
\nonumber \\
r_I(x,t)=\mathcal{I}_c\cos(\omega t - kx)+\mathcal{I}_s\sin(\omega t-kx).
\label{eq:soluitions2}
\end{eqnarray} 
In contrast to the in-phase response of the system to purely spatial stimuli,  the response to dynamic stimuli is characterized by  temporal delays relative to the responses to purely spatial stimuli.~This means that responses to spatiotemporal stimuli have both sine and cosine components, even when the stimulus only contains cosine components.  
By substituting (\ref{eq:soluitions2}) in (\ref{eq:pde}) with $\beta=0$ and $\gamma = 0$, we find the following solutions for the amplitudes of activation of the excitatory ($\mathcal{E}_c$ and $\mathcal{E}_s$) and inhibitory ($\mathcal{I}_c$ and $\mathcal{I}_s$) cells: 
\begin{eqnarray}
\mathcal{E}_c=j_0\frac{P_{ec}}{H}, \nonumber \quad 
\mathcal{E}_s=j_0\omega\frac{P_{es}}{H}, \nonumber \\
\mathcal{I}_c=j_0\frac{P_{ic}}{H}, \quad 
\mathcal{I}_s=j_0\omega\frac{P_{is}}{H},
\label{eq_amplitudes}
\end{eqnarray}  
where 
\[P_{ec} = \mu[\Psi_I+\Phi_I k^2][(k^2-b)^2+d]
+\omega^2k^2(\alpha D_{EE}-(1-\alpha) D_{EI})+\omega^2((1-\alpha) K_{EI}-\alpha(K_{EE}-1)),\]
\[P_{es} = \alpha((K_{IE}-k^2D_{IE})(k^2D_{EI}-K_{EI})-(k^2D_{II}-K_{II}-1)^2+\omega^2)+(1-\alpha)(k^2D_{EI}-K_{EI})(\Delta K+
k^2\Delta D),\]
\[P_{ic} = \mu[\Psi_E+\Phi_Ek^2][(k^2-b)^2+d]+\omega^2k^2(\alpha D_{IE}-(1-\alpha)D_{II}))+\omega^2((1-\alpha)(K_{II}+1)-\alpha K_{IE}),\]
\[P_{is} = (1-\alpha)((K_{EE}-1-k^2D_{EE})^2+(k^2D_{EI}-K_{EI})(K_{IE}-k^2D_{IE})+\omega^2)+\alpha(K_{IE}-k^2D_{IE})
(\Delta K+\Delta D k^2),\]
\noindent 
and where the denominator of every solution is the same:
\[H = \mu^2[(k^2-b)^2+d]^2+\omega^2[\left(\Delta D^2-2\mu\right)k^4+(4\mu b+2\Delta K \Delta D)k^2]+
\omega^2[\Delta K^2-2\mu(b^2+d)]+\omega^4,\]
\noindent 
with $\Delta D= D_{EE}-D_{II}$ and $\Delta K=K_{II}-K_{EE}+2$. 
In spite of its complexity, equation (\ref{eq_amplitudes}) can be plotted numerically, as we did in panels A and C of Figure~\ref{fig:sf_tf}.
The plots reveal a family of the familiar tuning functions.

As before, we find the intrinsic spatial frequency of the system by looking for the system's {\it maximum} response. 
The latter is expected where the denominator approaches its {\it minimum}. 
Since our goal is to understand how the maximum is controlled by temporal frequency $\omega$, in the following  
we focus on the analysis of the denominator in different regions of the parameters space. 

To begin, notice that the first term of the denominator~$H$ depends exclusively on spatial frequencies, the last term depends exclusively on temporal frequencies, while the middle term depends on the interaction of spatial and temporal frequencies. 
This structure suggests that the system can be studied using three distinct approaches:\label{approaches} 
\begin{description}
\item 
(1)  finding the {\it natural spatial frequency} of the system (i.e., its maximum response) under variation of spatial frequency $k$, defined by equation $\partial H/\partial k=0$ for a fixed temporal frequency $\omega$,

\item 
(2) finding the {\it natural temporal  frequency} of the system under variation of temporal frequency $\omega$, defined by equation $\partial H/\partial\omega=0$  for a fixed spatial frequency $k$, and

\item 
(3)  finding the {\it natural tuning speed} of the system for certain combination of $k$ and $\omega$, and under joint variation of  $k$ and $\omega$,  defined by equation $\partial H/\partial k=\partial H/\partial \omega=0$. 
\end{description}
Here we pursue first two of these approaches, in the next two sections.

\subsection{Variation of spatial frequency}

Taking the derivative $\partial H/\partial k \,| \, k_{res}(\omega)=0$ results in the following equation:
\begin{equation}
\omega^2=\frac{4\mu^2\left[(k_{res}^2-b)^2+d\right](k_{res}^2-b)}{2k_{res}^2(2\mu-\Delta D^2)-4\mu b-2\Delta D\Delta K}. 
\label{freq-den}
\end{equation}  
Notice that for $\omega=0$ the above equation reduces to  \[k_{\rm res}^2(0)=b,\]
reproducing our previous result for purely spatial stimuli ($\omega=0$) in equation~(\ref{eq:summary}) with the nonlinear terms removed.  

For the sake of understanding how stimulus temporal frequency ($\omega$) affects resonant spatial frequency ($k_{res}$)   of the neural chain, it is useful to write $b$ as $k_{res}^2(\omega = 0)$, which we represent by $k_{res}^2(0)$. 
For nonzero $\omega$, we rewrite equation (\ref{freq-den}) by approximating $k_{res}(\omega) \approx k_{res}(0)$ in the denominator of~(\ref{freq-den}) so we can study the influence of $\omega$ of $k_{res}$:
\begin{equation}
\omega^2=-\frac{2\mu^2\left[(k_{res}^2-k_{res}^2(0))+d\right] [k_{res}^2-k_{res}^2(0)] }{\Delta D [k_{res}^2(0)+\Delta K] }.
\label{eq:temporal_resonance}
\end{equation}
Given that $\omega^2$ is positive, the sign of the denominator of (\ref{eq:temporal_resonance}) will determine 
how $\omega$ affects the system behavior. 
In other words, varying  $\omega$ can change $k_{res}$ two ways, implying two \emph{spatial regimes}: 
\begin{enumerate}
\item[{[S1]}] 
When the denominator of (\ref{eq:temporal_resonance}) is negative, the numerator must be positive, implying that  
$k_{res}(\omega)^2-k_{res}^2(0) > 0$. 
Here $k_{res}$ increases with $\omega$, as shown is Figure~\ref{fig:sf_tf}A, approaching an asymptote where the denominator is zero.  
This regime is represented in Figure~\ref{fig:sf_tf}B by the red curve that approaches the asymptote (here represented by the vertical dashed line) from the left side, which is the side of low~$k_{res}$. 

\item[{[S2]}] 
When the denominator of (\ref{eq:temporal_resonance}) is positive, the numerator must be negative, implying that $k_{res}(\omega)^2-k_{res}^2(0) < 0$. 
In this regime, $k_{res}$ decreases with $\omega$  as shown in Figure~\ref{fig:sf_tf}C.
This regime is represented in Figure~\ref{fig:sf_tf}B by the black curve, which approaches the same asymptote as above, yet from the right side, which is the side of high~$k_{res}$.

\end{enumerate}
We know from our previous analysis (\ref{eq:summary}-\ref{eq:summary_rewrite}), that the shift of tuning to higher spatial frequencies is observed at higher stimulus contrasts. 
It is therefore reasonable to expect that regime~S2 should be found at higher stimulus contrasts.

This prediction has been recently corroborated in a physiological study of neuronal selectivity in the Middle Temporal (MT) cortical area of alert macaque monkeys. 
\citet{Pawar_etal_2017SfN,Pawar_etal_2018cosyne}  found that spatial frequency tuning of MT neurons  depends on the luminance contrast and temporal frequency of stimulation.~As contrast increased, preferences of cortical neurons shifted towards lower or higher spatial frequencies, more often the latter than the former. The changes in spatial frequency preference were largest at low temporal frequency, and they decreased as temporal frequency increased.
Notably, preferred spatial frequencies (1)~increased with temporal frequency at low stimulus contrast, and (2)~decreased with temporal frequency at high stimulus contrast, in agreement with the pattern of predictions summarized in~Figure~\ref{fig:sf_tf}B (Pawar et al., in preparation).

\subsection{Variation of temporal frequency}

Next consider the case when temporal frequency varies and spatial frequency does not. 
Here the maximum of $r_E$ is reach at a certain ``natural" or ``intrinsic" temporal frequency of the system. 
We estimate the natural temporal frequency using the condition 
\[\partial H/\partial\omega|_{\omega=\omega_{\rm res}}=0\] and obtain:
\vspace{3pt}
{\small
\begin{equation}
\omega_{\rm res}=\frac{1}{\sqrt{2}}\left(2\mu(b^2+d)-\Delta K^2-(\Delta D^2-2\mu)k^4-(4\mu b+2\Delta K\Delta D)k^2\right)^\frac{1}{2}.
\label{eq:w_res}
\end{equation}  } 
\noindent Each of the three terms on the right side of  (\ref{eq:w_res}) can be positive or negative,  as described in Table~\ref{tab:fixed_SF}, producing qualitatively different patterns of behavior described in the table. 

\begin{table*}[b!]\index{typefaces}
	\small
	\begin{tabular}{ | l | l | l | l | p{9cm} |}
		\hline
		& $2\mu(b^2+d)-\Delta K^2$ &  $\Delta D^2-2\mu$ & $4\mu b+2\Delta K\Delta D$ &  \\ \hline
		{\footnotesize T}1 & positive & negative & negative & 
		The maximum of $r_E$ occurs at a non-zero temporal frequency $\omega$ even at zero spatial frequency $k$, i.e., $\omega_{res}(0) \neq0$. 
		In this regime, $\omega_{res}$ monotonically increases as $k$ increases. \\ \hline
		{\footnotesize T}2 & positive & negative & positive & 
		Here $\omega_{res}(0) \neq0$ as above, The initially positive $\omega_{\rm res}$ decreases, so the maximum of $r_E(\omega)$ can disappear, within an interval of~$k$. At larger~$k$, $\omega_{\rm res}$ increases with~$k$.  \\  \hline
		{\footnotesize T}3 & positive & positive & negative & 
		Again, $\omega_{res}(0) \neq0$, but $\omega_{\rm res}$ increases and reaches a maximum, after which 
		it drops to zero at a certain $k$. No maximum of $r_E(\omega)$ occurs at larger $k$.    \\  \hline
		{\footnotesize T}4 & positive & positive & positive &  
		As above,  $\omega_{res}(0) \neq0$, yet here $\omega_{\rm res}$ decreases and disappear at a certain $k$. 
		No maximum of $r_E(\omega)$ occurs at larger $k$.   \\ \hline
		{\footnotesize T}5 & negative & negative & irrelevant &  
		No resonance occurs at $k=0$, but it appears at some~$k \neq 0$ when $\omega_{\rm res}=0$, and then increases with increasing~$k$.   \\ \hline
		{\footnotesize T}6 & negative & positive & negative &  
		No resonance occurs at $k=0$, but it can appear within a certain interval of $k$. No maximum of $r_E(\omega)$ occurs at larger $k$.    \\ \hline
		{\footnotesize T}7 & negative & positive & positive &   No maximum of $r_E(\omega)$ occurs 
		at any $k$.    \\
		\hline
	\end{tabular}
	\caption[][-18pt]{Regimes of intrinsic temporal frequency ($\omega_{res}$) of the system discovered by varying temporal frequency $\omega$ for a fixed spatial frequency~$k$ of the stimulus (using the second approach listed on page~\pageref{approaches}).}
	\label{tab:fixed_SF}
	\setfloatalignment{b}
\end{table*}

\begin{figure}[t!]
	\centering
	\includegraphics[width=1.00\linewidth]{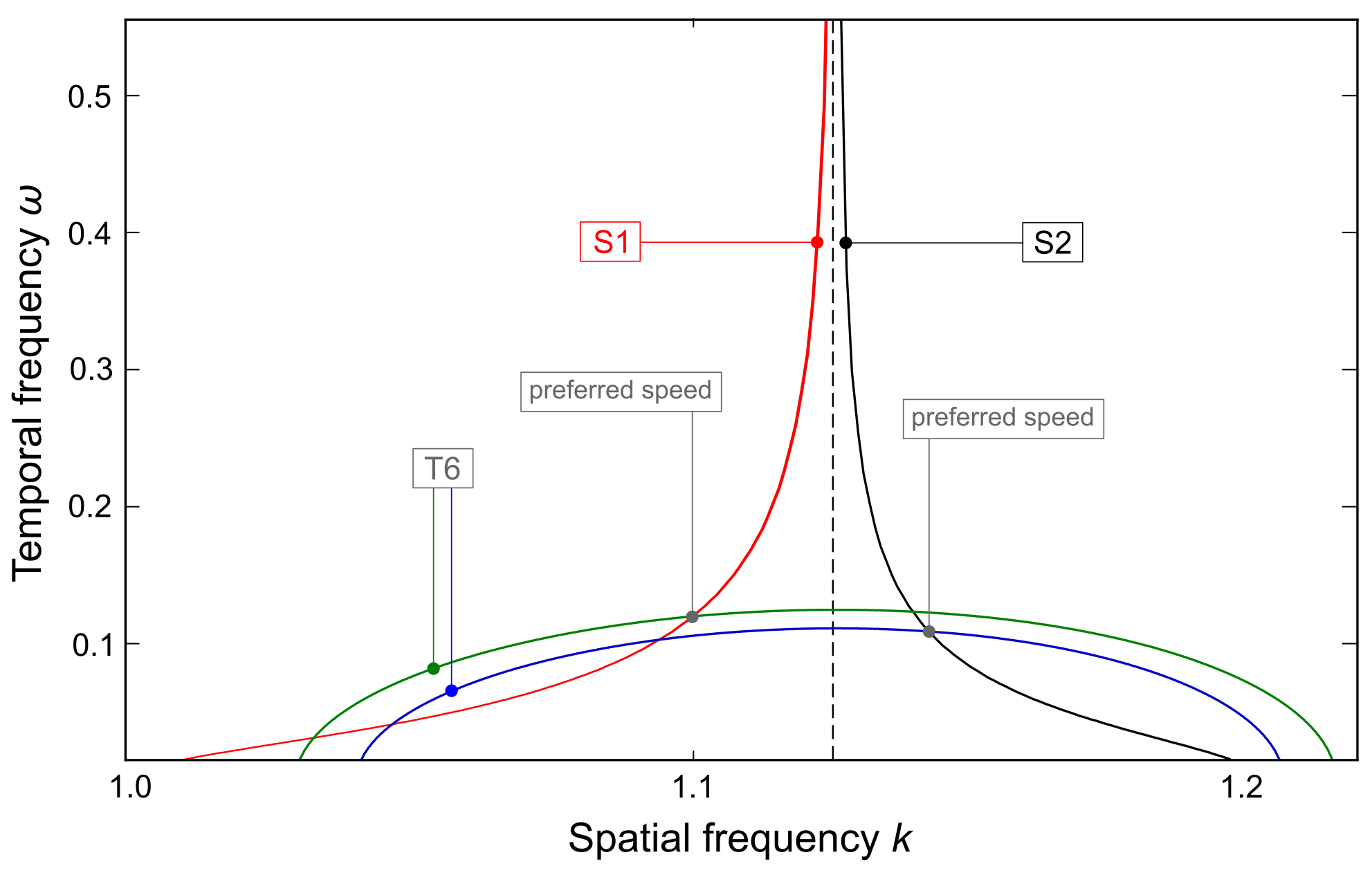}
	\caption[Intrinsic temporal frequency depends on stimulus spatial frequency][0pt]{
		The intrinsic temporal frequency~$\omega_{res}$  of the system depends on stimulus spatial frequency~$k$.
		Here $\omega_{res}$ is plotted as a function on $k$ in green and blue,  
		together with the two spatial resonance curves $k_{res}$ in red and black.~The latter two curves are reproduced from Figure~\ref{fig:sf_tf}B. 
		See text for detail. 
		\vspace{.1in}
		\newline
	} 
	\label{fig:fixed_SF}
\end{figure}

\noindent
The conditions listed in row~\textsc{\small T6}  correspond to the conditions used to predict the interaction of $k_{res}$ and $\omega$ illustrated in Figure~\ref{fig:sf_tf}. A key property of this prediction is reproduced in Figure~\ref{fig:fixed_SF} (red and black curves) together with the prediction derived from row~\textsc{\small T6} (green and blue curves):
\begin{description}
	\item[$\diamond$] The red and black curves were obtained using the first approach described on page~\pageref{approaches}, which is finding the natural  spatial frequency of the system~$k_{res}$ 
	for a fixed temporal frequency~$\omega$. 
	\item[$\diamond$] The green and blue curves were obtained using the second approach described on page~\pageref{approaches}, which is finding the natural  temporal frequency of the system~$\omega_{res}$ 
	for a fixed spatial frequency~$k$. 
\end{description}
\noindent 
Of these results, the red and green curves correspond to low stimulus contrasts while the black and blue curves correspond to high stimulus contrasts.~Accordingly, the intersection of the red and green curves, and also the intersection of the black and blue curves, correspond to the points of {\it preferred stimulus speed} of the system.

\section{Discussion}

We developed a model of spatially distributed cortical computation based on the spatially localized excitatory-inhibitory circuit originally proposed by Wilson \& Cowan (Figure~1A). 
In the distributed model, the localized circuits serve as "nodes" in a chain with the neighboring excitatory (E) and inhibitory (I) cells related through E-E, E-I, I-I, and I-E connections, i.e., forming a fully connected network with nearest-neighbor coupling (Figure~1B).

\vspace{.1in} \noindent \textbf{Intrinsic stimulus preference.}~We investigated response properties of this distributed system by first modeling its response to a "point" stimulus that activates a single node in the network. 
The point activation propagates through the chain of nodes and forms a steady-state spatial pattern distributed across the chain. 
A typical steady-state pattern has the form of a spatially damped oscillation~(Figure~2).  
The spatial frequency of this oscillation is  the natural or {\it intrinsic} spatial  frequency of the distributed network, determined by weights of excitatory and inhibitory connections between the cells within and between the nodes.  

\vspace{.1in} \noindent \textbf{Neural interference.}~Network response to a spatially distributed stimulus can be understood in terms of spatial interference between the point-processes elicited by the stimulus on multiple nodes: 
At {\it low stimulus contrasts}, where system behavior is linear, the interference amounts to superposition of the point-generated waves. 
Spatially periodic stimuli elicits maximum response when the spatial frequency of the stimulus matches the intrinsic spatial frequency of the network.   
And at {\it high stimulus contrasts}, where system behavior is nonlinear, predicting system response requires more complex analysis.  We developed an analytical solution to this problem  (Equation~9) allowing one to solve the system iteratively and determine its stimulus preferences at any contrast. 

\vspace{.1in} \noindent \textbf{Dynamics of stimulus preference.}~This analysis revealed that the preferred spatial frequency $k_{res}$ of the network should change as stimulus contrast increases, towards higher or lower frequency than at low contrasts.
The {\it direction} of change depends on the balance of 
excitatory or inhibitory processes in the network (Equation~\ref{eq:summary_rewrite}). 
In particular, $k_{res}$ is predicted to  increase with contrast in the network dominated by excitation, and to decrease with contrast in the network dominated by inhibition~(Figure~3).

\vspace{.05in} 
\noindent
The {\it amount } of change of $k_{res}$ is predicted to depend on stimulus temporal frequency $\omega$. 
At low stimulus contrast, $k_{res}$ increases with $\omega$, and at high stimulus contrast, $k_{res}$ decreases with $\omega$~(Figure~4). 
The latter result can be described in terms of $k_{res}$ at low and high stimulus contrasts converging at a certain "attractor" value of $k_{res}$ as $\omega$ increases. 
The "attractor" value is represented in Figure~4 by the vertical asymptote. 

\vspace{.05in} 
\noindent
Remarkably, the dynamics of neural preferences similar to that predicted by our model (Figure~4) was discovered recently in a  study of neuronal preferences in the cortical area MT of alert macaque monkeys (\citealt{Pawar_etal_2017SfN,Pawar_etal_2018cosyne}; Pawar et al., in preparation).

\newpage

\end{document}